\documentclass[11pt]{article}

\newcommand{\rnc}{\renewcommand}

\makeatletter \@addtoreset{equation}{section}
\rnc{\theequation}{\thesection.\arabic{equation}} \makeatother

\def\sqr#1#2{{\vcenter{\vbox{\hrule height.#2pt
\hbox{\vrule width.#2pt height#1pt \kern#1pt \vrule width.#2pt}\hrule
height.#2pt}}}}
\def\square{\mathchoice\sqr45\sqr45\sqr{2.1}3\sqr{1.5}3}

\def\a{\alpha}
\def\b{\beta}

\def\C{\Gamma}
\def\d{\delta}\def\D{\Delta}
\def\e{\epsilon}\def\ve{\varepsilon}

\def\f{\phi}\def\F{\Phi}\def\vf{\varphi}
\def\bphi{{\bar\phi}}
\def\h{\eta}
\def\k{\kappa}
\def\l{\lambda}\def\L{\Lambda}
\def\m{\mu}
\def\n{\nu}
\def\r{\rho}
\def\s{\sigma}\def\S{\Sigma}
\def\t{\tau}

\def\X{\Xeta}
\def\be{\begin{equation}}\def\ee{\end{equation}}
\def\bea{\begin{eqnarray}}\def\eea{\end{eqnarray}}
\def\ba{\begin{array}}\def\ea{\end{array}}
\def\x{\xi}
\def\o{\omega}
\def\p{\partial}

\let\la=\label
\let\bl=\bigl \let\br=\bigr
\let\Br=\Bigr \let\Bl=\Bigl
\let\bm=\bibitem

\def\ra{\rightarrow}
\def\nn{\nonumber}
\def\bd{\begin{document}}
\def\ed{\end{document}}
\def\ft#1#2{{\textstyle{{\scriptstyle #1}\over {\scriptstyle #2}}}}
\def\fft#1#2{{#1 \over #2}}
\newcommand{\eq}[1]{(\ref{#1})}
\def\eqs#1#2{(\ref{#1}-\ref{#2})}
\def\det{{\rm det\,}}
\def\tr{{\rm tr}}
\def\Tr{{\rm Tr}}
\newcommand{\ho}[1]{$\, ^{#1}$}
\newcommand{\hoch}[1]{$\, ^{#1}$}
\def\fdf{\phi^\dagger\phi}
\def\ffd{\phi\phi^\dagger}
\def\qq{\quad\quad}
\newcommand{\w}[1]{\\[0.#1cm]}
\def\lra{\leftrightarrow}

\def\cramp{\medmuskip = 2mu plus 1mu minus 2mu}
\def\cramper{\medmuskip = 2mu plus 1mu minus 2mu}
\def\crampest{\medmuskip = 1mu plus 1mu minus 1mu}
\def\uncramp{\medmuskip = 4mu plus 2mu minus 4mu}
\def\ben{\begin{equation}}
\def\een{\end{equation}}
\def\half{{\textstyle{1\over2}}}
\let\a=\alpha \let\b=\beta \let\g=\gamma \let\d=\delta \let\e=\epsilon
\let\z=\zeta \let\h=\eta \let\q=\theta \let\i=\iota \let\k=\kappa
\let\l=\lambda \let\m=\mu \let\n=\nu \let\x=\xi
\let\r=\rho
\let\s=\sigma \let\t=\tau \let\f=\phi  \let\y=\psi
\let\D=\Delta \let\Q=\Theta \let\L=\Lambda
\let\X=\Xi \let\P=\Pi \let\S=\Sigma \let\U=\Upsilon \let\F=\Phi \let\Y=\Psi
\let\W=\Omega
\let\la=\label \let\ci=\cite \let\re=\ref
\let\se=\section \let\sse=\subsection \let\ssse=\subsubsection
\def\nn{\nonumber} \def\bd{\begin{document}} \def\ed{\end{document}}
\def\ds{\documentstyle} \let\fr=\frac \let\bl=\bigl \let\br=\bigr
\let\Br=\Bigr \let\Bl=\Bigl
\let\bm=\bibitem
\let\na=\nabla
\let\pa=\partial \let\ov=\overline
\def\ba{\begin{array}}
\def\ea{\end{array}}
\def\ft#1#2{{\textstyle{{\scriptstyle #1}\over {\scriptstyle #2}}}}
\def\fft#1#2{{#1 \over #2}}
\def\del{\partial}
\def\vp{\varphi}
\def\sst#1{{\scriptscriptstyle #1}}
\def\oneone{\rlap 1\mkern4mu{\rm l}}
\def\td{\tilde}
\def\ie{\rm i.e.\ }
\def\dalemb#1#2{{\vbox{\hrule height .#2pt
        \hbox{\vrule width.#2pt height#1pt \kern#1pt
                \vrule width.#2pt}
        \hrule height.#2pt}}}
\def\square{\mathord{\dalemb{6.8}{7}\hbox{\hskip1pt}}}
\newcommand{\ap}{\alpha^\prime}
\newcommand{\bp}{\tilde \beta^\prime}
\def\0{{\sst{(0)}}}
\def\1{{\sst{(1)}}}
\def\2{{\sst{(2)}}}
\def\3{{\sst{(3)}}}
\def\4{{\sst{(4)}}}
\def\5{{\sst{(5)}}}
\def\6{{\sst{(6)}}}
\def\7{{\sst{(7)}}}
\def\8{{\sst{(8)}}}
\def\n{{\sst{(n)}}}
\def\cA{{{\cal A}}}
\def\cF{{{\cal F}}}
\def\tV{\widetilde V}
\def\tW{\widetilde W}
\def\tH{\widetilde H}
\def\tE{\widetilde E}
\def\tF{\widetilde F}
\def\tA{\widetilde A}
\def\im{{{\rm i}}}
\def\tY{{{\wtd Y}}}
\def\ep{{\epsilon}}
\def\vep{{\varepsilon}}
\def\R{\rlap{\rm I}\mkern3mu{\rm R}}
\def\bD{{{\bar D}}}

\def\R{\rlap{\rm I}\mkern3mu{\rm R}}
\def\bD{{{\bar D}}}
\def\R{{{\Bbb R}}}
\def\CN{{{\Bbb C}}}
\def\H{{{\Bbb H}}}
\def\CP{{{\Bbb C}{\Bbb P}}}
\def\RP{{{\Bbb R}{\Bbb P}}}
\def\Z{{{\Bbb Z}}}
\def\bA{{{\Bbb A}}}
\def\bB{{{\Bbb B}}}
\def\bC{{{\Bbb C}}}
\def\bR{{{\Bbb R}}}
\def\bD{{{\Bbb D}}}
\def\bE{{{\Bbb E}}}
\def\bZ{{{\Bbb Z}}}
\def\Re{{{\frak{Re}}}}
\def\Im{{{\frak{Im}}}}
\def\cosec{{\,\hbox{cosec}\,}}
\def\Gm{{\Gamma_{\!\! -}}}
\def\Gp{{\Gamma_{\!\! +}}}
\def\stan{{standard }}
\def\nonstan{{supernumerary }}

\def\cosech{{\hbox{cosech}}}

\def\etcyc{{\hbox{and cyclic}}}
\def\btheta{{\bar\theta}}

\def\bphi{\overline{\phi}}

\newcommand{\tamphys}{\it\small George P. and Cynthia W. Mitchell Institute for
Fundamental Physics\\
Texas A\&M University, College Station, TX 77843-4242}

\newcommand{\ictp}{\it\small International Center for Theoretical
Physics, 34100 Trieste, Italy}

\newcommand{\auth}{\large  S. Randjbar-Daemi\hoch{1} and
E. Sezgin\hoch{2}}

\begin{document}

\hfill{IC/2004/6}

\hfill{MIFP-03-25}

\hfill{hep-th/0402217}

\vspace{20pt}

\begin{center}


{\Large\bf Scalar Potential and Dyonic Strings in 6D Gauged Supergravity}


\vspace{30pt}

\auth

\vspace{15pt}

\begin{itemize}
\item [$^1$] \ictp \item [$^2$] \tamphys
\end{itemize}

\vspace{60pt}

{\sc Abstract}

\end{center}

In this paper we first give a simple parametrization of the scalar coset
manifold of the only known anomaly free chiral gauged supergravity in six
dimensions in the absence of linear multiplets, namely gauged minimal
supergravity coupled to a tensor multiplet, $E_6\times E_7\times U(1)_R$
Yang-Mills multiplets and suitable number of hypermultiplets. We then
construct the potential for the scalars and show that it has a unique
minimum at the origin. We also construct a new BPS dyonic string solution
in which $U(1)_R\times U(1)$ gauge fields, in addition to the metric,
dilaton and the 2-form potential, assume nontrivial configurations in any
$U(1)_R$ gauged $6D$ minimal supergravity coupled to a tensor multiplet
with gauge symmetry $G\supseteq U(1)$. The solution preserves $1/4$ of
the $6D$ supersymmetries and can be trivially embedded in the anomaly
free model, in which case the $U(1)$ activated in our solution resides in
$E_7$.

\vskip 10pt

\pagebreak \setcounter{page}{1}



\section{Introduction}


The most symmetric ground state solutions in all of the higher
dimensional supergravity theories, as the low energy limit of the
superstring theories or the M theory, are the flat 10 dimensional
manifolds ( $R^d\times flat\quad torus$) and the pp waves. Such
backgrounds have 32 real supersymmetries and are not a suitable starting
point for building realistic phenomenology. The Calabi-Yau
compactifications have less supersymmetries but also a lesser degree of
uniqueness. There exist many of them with a lot of moduli with unknown
potentials.

 In the gauged minimal supergravity theories in $D=6$ this is not the case.
 In fact in these theories the 6-dimensional flat spaces  do not solve
 the supergravity equations. The most symmetric solution is $R^4\times
 S^2$ \cite{Salam:1984cj} which  has been shown recently to be the
 $unique$ maximally symmetric solution of such models
 \cite{Gibbons:2003di}. This result has been obtained essentially in the
 minimal version of such models for which the gauge group is simply
 $U(1)$, i.e. the $D=6$ supersymmetric Einstein Maxwell theory, known as
 the Salam-Sezgin model in the literature \cite{Salam:1984cj}. The model
 by itself is anomalous but it can be embedded into an anomaly-free model
 \cite{Randjbar-Daemi:wc} with suitable Yang-Mills and hypermatter sector
 couplings \cite{{Nishino:dc}, {Nishino:1997ff}}.

The uniqueness of the supersymmetric $R^4\times S^2$ solution should be
contrasted with the plethora of solutions of Calabi-Yau type or the ones
with exceptional holonomy groups such as $G_2$ in higher dimensional low
energy string theories.  For this reason we consider this property as
very interesting and believe that such models deserve further study. This
is the aim of the present note.

This paper contains two results. First, after giving the multiplet
structure of general gauged $N=1$ supergravity models in $D=6$, we shall
concentrate on the hypermatter sector. This sector contains the scalars
and fermions which can be in anomaly free representation of the
Yang-Mills gauge groups and therefore are of primary importance in any
phenomenological application of such models. So far we know of only one
anomaly free gauged minimal supergravity in $D=6$
\cite{Randjbar-Daemi:wc} in the absence of linear multiplets
\footnote{The linear multiplet is a hypermultiplet in which one of the
four scalars is dualized to a 4-form potential. In this case, an
additional Green-Schwarz counterterm is possible for anomaly cancellation
\cite{Berkooz:1996iz}, and this may lead to new anomaly free models.}.
For this reason we shall construct in detail the scalar manifold and the
potential for the scalars in this particular model. However, our simple
parametrization of the quaternionic scalar manifold and the construction
of the scalar potential should be applicable in other cases too. The
first result of this construction is the observation that the scalar
potential admits a unique minimum at the origin. There are no moduli. As
we shall see this trivially implies the uniqueness of the $R^4\times S^2$
solution in the anomaly free full fledged model.

The second main point of the paper is the construction of new dyonic
string solutions. These solutions have a rather nontrivial structure and
leave $1/4$ of the $D=6$ supersymmetries unbroken. The search for such
solutions is motivated by the general philosophy that they can help us to
study string and field theories from a non-perturbative, semiclassical
point of view.

The choice of the model  to be considered here is dictated by anomaly
cancellation. It belongs to a general class of (1,0) supergravity models
constructed some time ago \cite{{Nishino:dc}, {Nishino:1997ff}}. It is
based on a six-dimensional $(1,0)$ supergravity theory coupled to a
tensor, Yang-Mills and hyper multiplets \cite{Randjbar-Daemi:wc}. At
present this is the only known gauged $(1,0)$ anomaly free supergravity
in $D=6$ in the absence of linear multiplets. Its string or M-theory
origin is still not quite well understood, although some progress has
been made in connecting a subset of the model to M-theory in
11-dimensions \cite{Cvetic:2003xr}, and Horava-Witten type construction
in 7-dimensions \cite{Gherghetta:2002nq}. Note, however, that in the
latter case the R-symmetry group is not gauged in the resulting $D=6$
supergravity.

Recently, our model has attracted interest in connection with  a
possibility of obtaining a small cosmological constant in $D=4$
\cite{Aghababaie:2003wz}.  These ideas make use of an extension of the
old magnetic monopole solution \cite{Randjbar-Daemi:1982hi} to a
situation in which 3-branes are distributed over a transverse $S^2$
\cite{{Carroll:2003db},{Navarro:2003vw}}.  In order for the mechanism to
work it is required that the supersymmetry breaking on the brane does not
propagate to the bulk. This is the weak point of the scheme for a finite
volume 2-manifold, as has been argued in a very general context in
\cite{Dvali:2002pe}. We hope that some other variations of the idea will
make a dent on this very important unsolved problem.

All the supersymmetric solutions of our model known so far, whether
compactifying solutions  with a direct product geometry or more involved
stringy solutions, use only the $U(1)_R$ gauge fields as part of their
ansatz. It seems quite difficult to excite the gauge fields in the
non-Abelian $E_6\times E_7$ component of the gauge group in search of a
supersymmetric solution. In this paper we report on one such solution in
which the gauge field in a $U(1)$ subgroup of $E_7$, in addition to that
of the $U(1)_R$ factor in the gauge group, is also nonzero. The
configuration is quite involved and does not have a 4-dimensional
Poincar\'e invariance. It has a natural dyonic stringy interpretation
similar to the one in \cite{Guven:2003uw}.
In fact it reduces to the solution in \cite{Guven:2003uw} upon setting
the $E_7$ gauge field to zero, which in turn generalizes the the $6D$
dyonic strings preserving $1/4$ supersymmetry found in
\cite{Duff:1995yh,Duff:1996cf}.

Thus, in our dyonic string solution, in addition to the tensor fields and
the gauge field of $U(1)_R$, a $U(1)$ gauge field embedded in $E_7$ as
well as the dilaton field are also active. The only fields which do not
participate in the solution are the $E_6$ gauge fields and the
hyperscalars.

The composition of the paper is as follows: In section 2 we give a
detailed description of the model with explicit form for the potential in
the hyperscalars. In this section we also give the field equations and
the supersymmetry transformation rules.  In section 3 we show that the
absolute minimum of the scalar potential is at $\phi=0$. This fact
excludes a solution of the form $M_4\times K_2$ with a nonzero vev  for
any of the gauge fields in $E_6\times E_7$ and with any unbroken susy. In
section 3 we discuss the ansatz for the dyonic string with a nonzero vev
for the gauge fields in $U(1)_R\times E_7$ as well as the tensor fields.
In section 4 we verify that our ansatz satisfies the field equations and
leaves $1/4$ of the original supersymmetries, i.e. one complex susy in
$1+1$ dimensions, unbroken. In the limit of a vanishing $E_7$ field it
reduces to the solution found earlier in \cite{Guven:2003uw} where only
$U(1)_R$ field was activated. In this section we also show that our
solution has a horizon at $r=0$, while at $r=\infty$ it approaches a cone
over a squashed $S^3\times {\rm Minkowski}_2$. The dilaton diverges in
both limits. Section 5 contains our conclusions. We give the anomaly
polynomial in an Appendix correcting the misprints of
\cite{Randjbar-Daemi:wc}.


\section{The Model}


\bigskip


\subsection{\bf Field Content and the Scalar Manifold}


\bigskip

The six-dimensional gauged supergravity model we shall study involves the
following $N=(1,0)$ supermultiplets \footnote{There also exists a linear
multiplet consisting of a 4-form potential, a symplectic Majorana-Weyl
spinor and three real scalars but it is not coupled in the model we study
here.}

\be \ba{llll} {\rm graviton} & e_M^r &\psi_{M+}^A &B_{MN}^{-} \w2 {\rm
tensor (dilaton)} &\varphi &\chi_{-}^A &B_{MN}^{+} \w2 {\rm hypermatter}
&\phi^\a &\psi_{-}^a \w2 {\rm Yang Mills} &A_{M} &\lambda_{+}^A \ea \ee

where coordinate basis and tangent space indices are denoted by $M,N,...$
and $r,s,...$, respectively. The antisymmetric tensor potentials, $
B_{MN}^{\pm}$, give rise to selfdual and anti-selfdual 3-form field
strengths. All the spinors are symplectic Majorana-Weyl, $A=1,2$ label
the doublet of the $R$ symmetry group $Sp(1)_R$ and $a=1,...,912$ labels
the $912$ dimensional pseudoreal representation of $E_7$. The chiralities
of the fermions are denoted by $\pm$.

The hyperscalars $\phi^\a,\ \a=1,..., 912 \times 2$ parameterize the
quaternionic Kahler coset

\be {Sp(456,1)\over Sp(456) \times Sp(1)_R} \la{coset} \ee

The global $Sp(456,1)$ symmetry is broken by the gauging of its compact
$E_7\times U(1)$ subgroup. The composite local $Sp(456)\times Sp(1)_R$
symmetry is left intact by this gauging. Thus, together with the
``external'' $E_6$ gauge symmetry, the full symmetry of the model is

\be [E_7\times E_6\times U(1)]_{\rm local} \times [Sp(456)\times
Sp(1)]_{\rm composite\ local} \ee

Note that $E_7$ in $Sp(456)$ is such that the $912$ of $Sp(456)$ is
irreducible under $E_7$. This spectrum is anomaly free
\cite{Randjbar-Daemi:wc} and so far is the only known anomaly free gauged
supergravity model in $D=6$ in the absence of linear multiplets.

It will prove useful to present the model of \cite{Randjbar-Daemi:wc} in
alternative forms. To this end, we need to introduce some notation and
outline the building blocks. To begin with, we define the vielbein,
$Sp(n)$ and $Sp(1)$ composite connections on the coset via the
Maurer-Cartan form as

\be (L^{-1} \p_\a L)^{aA}=V_\a^{aA}\ ,\quad\quad (L^{-1} \p_\a
L)^{ab}=A_\a^{ab}\ ,\quad\quad (L^{-1} \p_\a L)^{AB}=A_\a^{AB}\ .\la{mc1}
\ee
The vielbeins obey the following relations

\be g_{\a\b}V^\a_{a A} V^\b_{bB}=\Omega_{ab}\epsilon_{AB}\ ,\quad\quad
V^\a_{aA} V^{\b aB} +\a \lra \b = g^{\a\b} \delta_A^B\ , \ee

where $g_{\a\b}$ is the metric on the coset and $\Omega$ and $\epsilon$
are the symplectic invariant antisymmetric matrices. Next, we define the
components of the $E_7\times U(1)_R$ gauged Maurer-Cartan form as

\be (L^{-1} D L)^{aA}= P^{aA}\ ,\quad\quad (L^{-1} D L)^{ab}=Q^{ab}\
,\quad\quad (L^{-1} DL)^{AB}=Q^{AB}\ ,\la{mc2} \ee
where
\be DL=\left(d-g_1A^3T^3-g_7A^IT^I\right)L\ .\ee
Here $A^3_M$ and $A^I_M\, (I=1,...,133)$ are the $U(1)_R\times E_7$ gauge
fields and $g_1$ and $g_7$ are the corresponding gauge coupling
constants. The following relations hold

\be P^{aA}=  ({\cal D}\phi^\a ) V_\a^{aA} \ ,\quad\quad Q^{ab}=({\cal
D}\phi^\a ) A_\a^{ab}\ ,\quad\quad Q^{AB}=({\cal D}\phi^\a) A_\a^{AB}-g_1
A^3 (T^3)^{AB}\ ,\ee
The covariant derivative can be written as

\be {\cal D}_M \phi^\a = \p_M \phi^\a - g_1 A^3_M K^{3\a} -g_7 A_M^I
K^{I\a}\ ,\ee
where $K^{3\a}$ and $K^{I\a}$ are the Killing vectors associated with
$E_7 \times U(1)_R \subset Sp(456,1)$ isometry.

Other building blocks to define the model are certain $C$-functions on
the coset \eq{coset}.  These were defined in \cite{Nishino:1997ff}, and
studied further in \cite{Percacci:1998ag}
 where it
was shown that they can be expressed as
\bea C^3_{AB} &=& \left( L^{-1} T^3 L\right)_{AB}\ ,\quad\quad C^I_{AB}\left( L^{-1} T^I L\right)_{AB}\ , \nn\w2 C_3^{aA}&=&\left( L^{-1} T^3
L\right)^{aA}\ ,\quad\quad C_I^{aA}=\left( L^{-1} T^I L\right)^{aA}\ ,
\la{cf} \eea
where $T^3$ and $T^I$ are the anti-hermitian generators of $U(1)_R$ and
$E_7$.


\subsection{ The Choice of L}


From the foregoing description it is clear that the main ingredient in
this construction is the section L, which maps the coset $G/H$ to the
group manifold $G$. We begin thus with a brief description of the groups
involved here. Firstly, by $Sp(n,1)$ what is really meant is the group of
pseudounitary (2n+2)-dimensional matrices

\be Sp(n+1)\, \cap \, SU(2n,2)\ee

It is convenient to represent these matrices by $(n+1)$-dimensional
arrays whose elements are 2-dimensional matrices,

\be
\begin{array}{lcll}
g &= &({g_{\mu}}^{\nu}) &\mu ,\nu \,=0,1,\ldots n\\
&= &({g_{\mu A}}^{\nu B}) &A,B=1,2
\end{array}
\ee

With this notation we have two kinds of metric:

\newcommand{\sigx}{{\sigma}_{2}}
\newcommand{\onex}{{1}_{2}}

\begin{eqnarray}
J &= &diag(\sigx,\ldots ,\sigx ,-\sigx )\nonumber\\
\eta &= &diag(\onex ,\ldots ,\onex ,-\onex ) \end{eqnarray}

where $\sigx$ and $\onex$ denote the matrices

$$ \sigx = \left( \begin{array}{cc} 0 &-i\\ i &0
\end{array} \right), \; \; \;
\onex = \left( \begin{array}{cc} 1 &0\\0 &1
\end{array} \right) $$

The group elements are required to satisfy two conditions

\begin{eqnarray}
g^{t}Jg &= J &(symplectic)\nonumber\\
g^{\dag} \eta g &= \eta &(pseudounitary) \end{eqnarray}

where $g^{t}$ and $g^{\dag}$ denote transpose and hermitian
conjugate---in the (2n+2)-dimensional sense--- respectively. It is
straightforward now to show that each $2 \times 2$ element in the
(n+1)-dimensional array satisfies the reality condition,

\begin{equation}
{g_{\mu}}^{\nu} = \sigx {g_{\mu}}^{\nu \dag t}\sigx \label{real}
\end{equation}

It can be interpreted as a real quaternion.

Having defined the group $USp(n,1)$ we may restrict to its maximal
compact subgroup, $USp(n) \times USp(1)$ by choosing

$$ {g_{0}}^{\mu}={g_{\mu}}^{0}=0,\,\,\mu=1,\ldots ,n $$

The matrices ${g_{0}}^{0}$ belong to SU(2), i.e.

$$ USp(1) = Sp(1) \cap SU(2) =SU(2)$$

To coordinatize the manifold, consider the `boost',

\begin{equation}
L_{\phi}= \left( \begin{array}{cc} a+b\phi \phi^{\dag} &\phi \\
\phi^{\dag} &c
\end{array} \right)
\end{equation}

where $\phi$ is a $2n \times 2$ matrix, $\phi^{\dag}$ is its hermitian
conjugate, $a,b$ are real and proportional to the $2n \times 2n$ identity
matrix, and $c$ is real and proportional to the $2\times 2$ identity
matrix. We can write

$$ \phi = \left( \begin{array}{l} \phi_{1}\\
\vdots \\ \phi_{n} \end{array} \right) $$

where the elements of this column are $2 \times 2$ matrices satisfying
the reality condition given above (and repeated below). It follows that
$L_{\phi}$ belongs to the group $USp(n,1)$ provided it is also unitary,

$$ L_{\phi}^{\dag} \eta L_{\phi} = \eta $$

This is achieved by choosing

\begin{equation}
\begin{array}{lll}
a &= &1\\
b &= &(-1+\sqrt{1+\phi^{\dag}\phi})/\phi^{\dag}\phi \\ c
&=&\sqrt{1+\phi^{\dag}\phi}
\end{array}
\end{equation}

In obtaining this result we have used the fact that the $2 \times 2$
matrix $\phi^{\dag}\phi$ is proportional to the identity $\onex$.

In our problem $n=456$ and thus the scalars have $912 \times 2$ complex
components. As real quaternions they can be considered  as a $456 \times
1$ matrix whose elements, $\phi _{m} $, are $2 \times 2$ matrices subject
to the reality condition

\begin{equation}
\phi _{m}^{*} = \sigma _{2}\, \phi _{m}\, \sigma _{2},\; \; m = 1, \ldots
,456
\end{equation}

where $\sigma _{2} $ is the Pauli matrix, then we have a total of $456
\times 4 = 1824$ real components. Thus, the following two notations are
equivalent:

\be \phi^a{}^A \ \ \ \leftrightarrow\ \ \ (\phi_m)_{A'}{}^A \ee

where $a=1,...,912$ and $A,A'=1,2$. Using the particular form of the
coset representative described above, the following $C$-functions take a
particularly simple form

\be C^3_{AB} =(1+ |\phi|^2)\,(T^3)_{AB}\ ,\quad\quad
C^I_{AB}=(\phi^\dagger T^I \phi)_{AB}\ ,\la{c12}\ee

where $ |\phi|^2 \equiv \tr\,\phi^\dagger\phi$. These are the only
components we need in order to construct the scalar potential.

\bigskip


\subsection{\bf Field Equations and Supersymmetry Transformation Rules}


\bigskip

The Lagrangian for the anomaly free model we are studying can be obtained
from \cite{Nishino:dc} or \cite{Nishino:1997ff}. We shall use the latter
in the absence of Lorentz Chern-Simons terms and Green-Schwarz anomaly
counterterms. Thus, the bosonic sector of the Lagrangian is given by
\cite{Nishino:1997ff}

\bea
{\cal L} &=& R\, {*\oneone} - \ft14 {*d\vf}\wedge d\vf - \ft12 e^\vf\,
*H\wedge H - \ft12 e^{\ft12\vf}\, \tr \left(*F_\wedge F\right)\nn\w2 &&
 - \ft12 {*{\cal D}\phi^\a}\wedge {\cal D}\phi^\b\,g_{\a\b} - 4 \,
e^{-\ft12\vf}\,(\tr\, C^2) \, {*\oneone}\ , \eea
where
\bea dH &=& \ft12 \tr\, F\wedge F \nn\w2
\tr \left(*F_\wedge F\right) &\equiv& *F^3\wedge F^3 +*F^I\wedge F^I
+*F^{I'}\wedge F^{I'}\ ,\quad I=1,...,133, \ \ I'=1,2,...,78\ ,\nn\w2
\tr\, C^2 &\equiv& g_1^2 C^3_{AB} C^{3,AB}+ g_7^2 C^I_{AB} C^{I,AB}\ .
\la{tc}\eea

We have let $A_\mu\ra A_\mu/{\sqrt v}$ and $g\ra g\sqrt v$ in the results
of \cite{Nishino:1997ff} to absorb the factors $v_1,v_6,v_7$ defined in
the appendix, so that that the normalizations of the Yang-Mills kinetic
terms and the Chern-Simons terms in $H$ are the same as those in
\cite{Nishino:dc}. The bosonic field equation following from the above
Lagrangian are \cite{Nishino:dc,Nishino:1997ff}

\bea
R_{MN} &=& \ft14 \del_M\vf\, \del_N\vf + \ft12 e^{\ft12\vf}\,
\tr\,(F^2_{MN} - \ft18 F^2\,g_{MN}) + \ft14 e^\vf\, (H^2_{MN} -
\ft16 H^2\, g_{MN}) \nn\w2
&& + \ft12 P_M^{aA} P_{NaA} + e^{-\ft12\vf}(\tr\, C^2)  \ g_{MN}\
, \nn\w2
\square\, \vf&=& \ft14 e^{\ft12\vf}\,\tr\, F^2 + \ft16 e^\vf\, H^2
-4\, e^{-\ft12\vf}\,\tr\, C^2 \nn\w2
d\big(e^{\ft12\vf}\,*F^3\big) &=& e^\vf\, {*H}\wedge F^3 -g_1 *P^{aA}
C^3_{aA}\ ,\nn\w2
d\big(e^{\ft12\vf}\,*F^I\big) &=& e^\vf\, {*H}\wedge F^I -g_7 *P^{aA}
C^I_{aA}\ ,\nn\w2
d\big(e^{\ft12\vf}\,*F^{I'}\big) &=& e^\vf\, {*H}\wedge F^{I'} \
,\nn\w2
d\left(e^\vf\, *H\right) &=& 0\ ,\nn\w2
{\cal D}_M P^{MaA}  &=& 2 e^\vf \left( g_1^2 C_3^{AB} (C_3)^a{}_B +g_7^2
C_I^{AB} (C_I)^a{}_B\right)\ . \la{e7} \eea

The local supersymmetry transformations of the fermions, up to cubic
fermion terms that will not effect our results for the Killing spinors,
are given by \cite{Nishino:1997ff}

\bea
\d \psi_M &=& {\cal D}_M \ve + \ft1{48} e^{\ft12\vf} H_{NPQ}^+\C^{NPQ}
\C_M\,\ve \ ,\nn\w2
\d\chi &=&\ft14\left( \C^M\p_M \vf  -\ft16 e^{\ft12\vf} H_{MNP}^-\C^{MNP}
\right)\ve\ , \nn\w2
\d \l_A^3 &=& -\ft18 F_{MN}^3\C^{MN}\ve_A
    -g_1 e^{-\ft12\vf} C^3_{AB}~\ve^B \ ,
\nn\w2
\d \l_A^I &=& -\ft18 F_{MN}^I\C^{MN}\ve_A
    -g_7 e^{-\ft12\vf} C^I_{AB}~\ve^B \ ,
  \nn\w2
\d \l_A^{I'} &=& -\ft18 F_{MN}^{I'}\C^{MN}\ve_A\ ,\nn\w2
\d\psi^a &=&  P_M^{a A} \C^M \ve_A \ , \nn\w2
\la{susy1} \eea

where ${\cal D}_M\ve_A = \p_M\ve_A +\ft14 \o_{Mrs}\C^{rs} \ve_A +
Q_{MA}{}^B \ve_B$. In addition to the constant re-scalings of $(A_\mu,g)$
mentioned above, we have also re-scaled $\l \ra \l / {\sqrt v}$ in the
results of \cite{Nishino:1997ff}. Furthermore, the transformation rules
for the gauge fermions differ from those in \cite{Nishino:dc}, and used
in \cite{Guven:2003uw}, by a field redefinition.

\bigskip


\subsection{\bf The Potential and its Minimum}


\bigskip

It is convenient to re-write the hyperscalar field equation as

\be g_{\a\b} {\cal D}_M{\cal D}^M \phi^\b = {\p V\over \p\phi^\a}\ . \ee

Upon the use of \eq{c12}, we obtain the potential

\bea V(\phi) &=& 4 \, e^{-\ft12\vf}\,(\tr\, C^2)\nn\w2
&=& e^{-\ft12\vf} \left[ 2g_1^2 (1+|\phi|^2)^2-g_7^2 \tr\,(\phi^\dagger
T^I\phi)^2\right]\ . \la{pot}\eea

Observe that since $T^I$ are anti-hermitian, the second term is positive
definite by itself, as is the first term. From the above potential, it is
obvious that the absolute minimum is at $\phi^\a =0$. Thus the potential
has a unique minimum. There are no moduli. Note that if we could set
$g_1=0$ there could be other nontrivial configurations which could break
$E_7$ spontaneously. However in this particular model $g_1$ has to be
different from zero for the anomaly cancellation. The nonvanishing of
$g_1$ is also the basic reason why the manifold $R^4\times S^2$ is a
solution. Essentially, at the minimum of the hyperscalars  the
exponential potential for the dilaton is given by $ 2g_1 ^2
e^{-\ft12\vf}$. For a constant dilaton this acts like a 6-dimensional
cosmological constant. When a $U(1)$ gauge field assumes a magnetic
monopole configuration on $S^2$ we obtain the solution $R^4\times S^2$.

By examining the susy transformation rules it becomes clear that there
can be no product space solution of the form $M_4\times K_2$ with a
nonzero vev of any of the non-Abelian gauge fields preserving any amounts
of supersymmetries.

The fact that the minimum of the hyperscalar potential is at $\phi=0$
implies that the $E_7$ symmetry can not be broken spontaneously by a vev
of the hyperscalars at the tree level. The only possibility of a tree
level breaking of $E_7$ (as well as $E_6$) is to give a vev to the
components of vector potential of these groups tangent to the internal
manifold $S^2$. If the monopole sits in $E_6\times E_7$ factor, the
configuration  is generally unstable, unless the monopole charge is
chosen to be the least possible value \cite{Randjbar-Daemi:1983bw}. Since
such configurations also break all the $D=6$ supersymmetries it follows
that at the tree level $E_6\times E_7$ and susy break at the same scale.

The mass of the fluctuations of  the scalar fields around the minimum at
$\phi-0$ will have two contributions for their masses. The first is the
mass term coming from the potential in \eq{pot} and the second is the KK
mass originating from the fact that the scalars are charged with respect
to the $U(1)\times E_7$ gauge fields. Therefore a magnetic monopole
background sitting in this group will generate a nonzero mass for all the
scalars, unless the magnetic charge of the two groups cancel out
mutually. If the effective magnetic coupling of a scalar field on $S^2$
is n, then the mass squared of the lightest Kaluza Klein mode in the
expansion of $\phi$ will be proportional to $|n|/a^2$ where $a$ is the
radius of $S^2$.

On the other hand since the fermions will couple to the magnetic monopole
embedded in $E_7$, there will be many chiral fermions in the low energy
spectrum in $R^4$, exactly in the same manner as in
\cite{Randjbar-Daemi:wc} where the monopole in a $U(1)$ subgroup of $E_6$
gave rise to two families of $16$ of $SO(10)$ in $R^4$. Several models of
this type in which the Higgs scalars may originate from the extra
components of the gauge field have been studied in detail in
\cite{Dvali:2001qr}

Having established that the minimum of the hyperscalar potential is at
$\phi=0$ the proof of the uniqueness of the $R^4\times S^2$ solution
should follow along the same lines as given in  \cite{Gibbons:2003di} for
the Salam-Sezgin model.

\bigskip


\section{The Ansatz, the $F$ and $H$  Field Equations and Supersymmetry}


\bigskip

In this section we present the dyonic string ansatz and determine
the equations that follow from the requirement of $F$ and $H$
field equations, and supersymmetry. Once these equations are
satisfied, we show that the Einstein and dilaton field equations
are automatically satisfied as well. We then proceed to solve all
the required  equations in section 4 where we present our dyonic
string solution.


\subsection*{\bf  The Ansatz }


\bigskip

Now we turn to the dyonic string ansatz. But before stating our ansatz we
will briefly summarize all the known brane solutions in our model. In
\cite{Gibbons:2003di} it was shown that the most general solution of our
model compatible with the Poincar\'e symmetry in $R^4$ is a 3-brane with
warped metric. The brane is a $\delta$-function singularity which can
also be interpreted as a deficit angle in the 2-dimensional transverse
space. This solution breaks all the supersymmetries. It reduces to 1/2
supersymmetric solution when the deficit angle vanishes.  In
\cite{Guven:2003uw} solutions of the type $AdS_3\times S_3$ as well
dyonic string solution  have been studied. It has also been shown that
the $AdS_3\times S^3$ solution goes over to the maximally symmetric
$R^4\times S^2$ configuration.

Our solution will be a generalization of the dyonic string of
\cite{Guven:2003uw}, in which in addition to the $U(1)_R$ gauge field a
$U(1)$ component in $E_7$ will also be nonzero, in a nontrivial way. We
thus start from the following ansatz:

\bea
ds_6 ^2 &=& c^2\, dx^\mu\, dx_\mu + a^2\, (\sigma_1^2 + \sigma_2^2) +
       b^2\, \sigma_3^2 + h^2\, dr^2\ ,\nn\w2
H &=& P\, \sigma_1\wedge \sigma_2\wedge\sigma_3 + u\, d^2x\wedge dr\ ,
\nn\w2
F^3 &=& k\, \sigma_1\wedge \sigma_2 ,\nn\w2
F^{7} &=& v\, \sigma_1\wedge \sigma_2 +v'\, \sigma_3\wedge dr\
,\la{an}\eea

where $d^2 x =dx^\mu\wedge dx^\nu \epsilon_{\mu\nu}$, $k$ is a constant,
$a$, $b$, $c$, $h$, $u$, $v$, $P$ and $\varphi$ are functions of $r$, and
the $\sigma_i$ are left-invariant 1-forms on the 3-sphere, satisfying the
exterior algebra
\be d\sigma_i = -\ft12 \ep_{ijk}\, \sigma_j\wedge \sigma_k\ . \ee
They can be represented, in terms of Euler angles
$(\theta,\varphi,\psi)$, by
\be \sigma_1 + \im\, \sigma_2 = e^{-\im\, \psi}\, (d\theta + \im\,
\sin\theta\, d\varphi)\,,\qquad \sigma_3 = d\psi +  \cos\theta\,
d\varphi\ . \ee

The function $h$ which may be removed by a coordinate transformation,
$dr'=h(r)\,dr$, will be chosen later to simplify the solution. Locally,
we can choose the potential for $F^3$ to be given by $A^3=-k\, \sigma_3$,
and for $F^7$ by $A^7=-v\, \sigma_3$. It is also useful to record

\bea *F^3 &=& {khbc^2\over a^2}\,\s_3\wedge dr\wedge d^2x\ ,\nn\w2
*F^7 &=& c^2\left({v'a^2\over hb}\,\s_1\wedge\s_2+{vhb\over
a^2}\,\s_3\wedge dr\right)\wedge d^2x\ ,\nn\w2
*H&=& -{uba^2\over hc^2}\,\s_1\wedge\s_2\wedge\s_3 -{Phc^2\over.
ba^2}\,d^2x\wedge dr\ . \eea

\bigskip


\subsection*{\bf The $F$ and $H$  Field Equations}


\bigskip

The $H$-field equation and the Bianchi identity $dH=\ft12 \tr\,F^2$ are
solved, respectively, by

\be u=-{Q_0hc^2 \over ba^2}\, e^{-\vf}\ , \quad\quad P=P_0-\ft12 v^2\
,\la{up} \ee

where $P_0$ and $-Q_0$ are integration constants. The $F^3$ and $F^7$
field equations are solved by

\be b^2= Pe^{\ft12\vf}\ ,\quad\quad\quad  v' = {v_0hb\over
a^2c^2}e^{-\ft12\vf} \ ,\la{bv} \ee
where $v_0$ is an integration constant.

\bigskip


\subsection*{\bf Killing Spinor Conditions}


\bigskip

Next, we examine the consequences of supersymmetry. Following
\cite{Guven:2003uw}, we impose the following conditions on the which
break supersymmetry by a factor of four:

\be \ft12 \Gamma^{12} \ve_A = (T^3)_A{}^B \ve_B\ ,\quad\quad
\C_{1234}\,\ve_A = \ve_A\ . \la{e12}\ee

Thus, the conditions $\delta\l^3=0$ and $\delta\l^7=0$ give

\be a^2 = {ke^{\ft12\vf}\over 2g_1} \ ,\quad\quad {v'\over v}= {hb\over
a^2} \ .\la{av}\ee

The condition $\d\psi^a=0$ is trivially satisfied, while $\d\chi=0$ is
solved by

\be \vf'=-e^{\ft12\vf} \left({u\over c^2}+{Ph\over ba^2}\right)\
.\la{fp}\ee

There remains the supersymmetry transformations of the gravitini. To this
end, it is useful to note that in the orthonormal frame defined by

$$ e^{\td 0} = c\, dt\ ,\quad e^{\td 1} = c\, dx\, \quad e^1= a\,
\sigma_1\, \quad e^2=a\, \sigma_2\, \quad e^3=b\, \sigma_3\ ,\quad
e^4=h\, dr\ , \la{frame}$$

the non-vanishing components of the spin connection take the form

\bea &&\omega_{23} = -\fft{b}{2a^2}\, e^1\ ,\quad \omega_{31} =
-\fft{b}{2 a^2}\, e^2\ ,\quad \omega_{12} = \left(\fft{b}{2a^2} -
\fft1{b}\right)\, e^3\ , \nn\w2 &&\omega_{14}= \fft{a'}{a\, h}\, e^1\
,\quad \omega_{24}= \fft{a'}{a\, h}\, e^2\ ,\quad \omega_{34}=
\fft{b'}{b\, h}\, e^3\ ,\quad \omega^\mu{}_4= \fft{c'}{c\, h}\, e^\mu\
.\la{omega} \eea

Using these results, and taking $\ve =\ve(r)$, it follows from
$\delta\psi_\mu=0$, $\d\psi_i=0\, (i=1,2)$ and $\d\psi_3=0$,
respectively, that

\bea {c'\over c} &=& \ft14 e^{\ft12\vf}\left({u\over c^2}-{Ph\over
ba^2}\right)\ ,\la{cp}\w2
{a'\over a} &=& \ft14 e^{\ft12\vf}\left(-{u\over c^2}+{Ph\over
ba^2}\right) -{bh\over 2a^2}\ ,\la{ap}\w2
{b'\over b} &=& \ft14 e^{\ft12\vf}\left(-{u\over c^2}+{Ph\over
ba^2}\right) +{bh\over 2a^2} +{(k g_1-1)h\over b}\ .\la{bp}
\eea
Finally, $\d\psi_4=0$ gives
\be {\ve}' = \ft18 e^{\ft12\vf}\left({u\over c^2}-{Ph\over ba^2}\right)\
. \la{eps}\ee

Comparing with \eq{cp}, we learn that

\be \ve(r)=c^{1/2}\,\ve_0\ . \ee

\bigskip


\subsection*{\bf The Einstein and Dilaton Field Equations}


\bigskip

Let us begin by writing the Einstein and dilaton equations given in
\eq{e7} as

\be R_{MN}=T_{MN}\ ,\quad\quad \square\, \vf=J\ .\ee

Substitution of the ansatz into these equations yields rather complicated
field equations which have provided in the Appendix. Instead of solving
these complicated second order field equations, it is much easier to show
that they are automatically satisfied once the Killing spinor conditions,
and the $F$ and $H$-field equations/Bianchi identities are satisfied. To
see this, let us first introduce the notation

\be \d \psi_M = {\widetilde D}_M \ve \ ,\quad\quad\quad \d\chi =\D \ve\
.\ee

It is then straightforward to show that

\bea && \C^N [{\widetilde D}_M,{\widetilde D}_N] \ve
\left(R_{MN}-T_{MN}\right)\,\C^N \ve + X_M \ , \nn\w3
&& \C^M[{\widetilde D}_M,\D]\ve = (\square\,\vf-J)\, \ve + Y \ , \eea

where $X_M$ and $Y$ are expressions which vanish upon the use of the $F$
and $H$ field equations/Bianchi identities. Therefore, the dilaton
equation is evidently satisfied, and so is the Einstein equation, once we
note that $R_{MN}$ is diagonal for our ansatz, as shown in the Appendix.

\bigskip


\section{\bf The Dyonic String Solution}


\bigskip

 As in  \cite{Guven:2003uw}, we make the gauge choice

\be h= -{2a^2bc^2\over r^3}\ . \la{gc}\ee

Then, defining the combinations

\be \vf_\pm= \vf \pm 4\ln c\ ,\ee

we find  from \eq{fp} and \eq{cp}, with the help of \eq{up},\eq{bv} and
\eq{av} that

\bea \vf_-' &=& {4Q_0\over r^3} e^{-\ft12\vf_-}\ ,\nn\w2
\vf_+'&=& {4P_0\over r^3}\left( e^{\ft12\vf_+}-\b^2
e^{-\ft12\vf_+}\right)\ ,\nn\w2
\b &=& {v_0\over \sqrt{2P_0}}\ ,\la{vm} \eea

These have solutions

\be e^{\vf} = \b H_1 \coth\,(\b H_2)\ ,\quad\quad  c^{-4}= {H_1\over \b}
\tanh\,(\b H_2)\ ,\la{s1} \ee

where

\be H_1={\widetilde Q}_0+{Q_0\over r^2}\ ,\quad\quad H_2={\widetilde
P}_0+{P_0\over r^2}\ .\la{h12} \ee

Here, we have introduced the integration constants ${\tilde Q}_0$ and
${\tilde P}_0$. Next, from \eq{bp}, making use of \eq{up}, \eq{bv}, and
\eq{av} we find

\be P_0= {k(1-kg_1)\over 2g_1}\ . \la{pk}\ee

Note that \eq{up}, \eq{bv}, and \eq{av} also yield the results

\be v = {v_0 e^{-\ft12\vf}\over c^2}\ ,\quad\quad P= {P_0\over \cosh^2\,
(\b H_2)}\ . \la{vp}\ee

The remaining quantities in the ansatz, namely, the functions $(a,b,u)$
can now be evaluated in terms of $(\vf,c)$ via algebraic equations
\eq{av}, \eq{bv}, \eq{up} and \eq{vp}. The result for the ansatz can now
be summarized as follows:

\bea
ds^2_6 &=& \sqrt{\b H_1 \coth(\b H_2)}\left[ {dx^\mu dx_\mu\over H_1}  +
{k\over 2g_1}\left(\s_1^2+\s_2^2\right)+{P_0\over \cosh^2(\b
H_2)}\,\s_3^2 +{P_0\b^2k^2\over g_1^2\sinh^2(\b H_2)}{dr2\over
r^6}\,\right] ,\nn\w2
e^{\vf} &=& \b H_1 \coth\,(\b H_2)\ ,\nn\w2
H&=& {P_0\over \cosh^2\,(\b H_2)}\,\s_1\wedge\s_2\wedge \s_3 -d^2x\wedge
d H_1^{-1}\ ,\nn\w2
F^3 &=& k\, \sigma_1\wedge \sigma_2 \ ,\nn\w2
F^7&=&\sqrt{2P_0}\left( \tanh\,(\b H_2)\,\s_1\wedge\s_2 +{\b\over
\cosh^2\,(\b H_2)}\,\s_3\wedge dH_2\right)\ ,\la{hff} \eea

with \eq{pk} holding, and $(H_1, H_2)$ and $\b$ are defined in \eq{h12}
and \eq{vm}, respectively, and $(P_0,k,v_0)$ are constants. Furthermore,
from \eq{bv} \eq{av},\eq{pk} and \eq{vp}, we learn that

\be P_0 \ge 0\ ,\quad\quad k  \le {1\over g_1}\ . \ee

In the limit of $\b \ra 0$, the $U(1)\subset E_7$ gauge field vanishes
and the non-vanishing fields become

\bea ds^2_6 &=& \sqrt{H_1\over H_2}\left[ {1\over H_1} dx^\mu dx_\mu +
{k\over 2g_1}\left(\s_1^2+\s_2^2\right)+P_0\,\s_3^2 +{P_0^2\over
g_1^2(H_2)^2}{dr^2\over r^6} \right] ,\nn\w2
e^\vf &=& {H_1\over H_2}\ ,\nn\w2
H&=& P_0\,\s_1\wedge\s_2\wedge \s_3 -d^2x\wedge d H_1^{-1}\ ,\nn\w2
F3 &=& k\, \sigma_1\wedge \sigma_2 \ ,\la{hff2} \eea

with \eq{pk} holding. This is the solution obtained in
\cite{Guven:2003uw}.

Turning to our solution \eq{hff}, in order to study its global structure,
following \cite{Guven:2003uw} we change to a new radial coordinate $\rho$
related to $r$ as

\be {\widetilde P}_0 +{P_0\over r^2} = {P_0^2\over \rho^4}\ , \ee

Our solution then is given by \eq{hff} with

\be H_1 = \left({\widetilde Q}_0-{Q_0{\widetilde P}_0 \over
P_0}\right)+{Q_0P_0\over \rho^4}\ ,\quad\quad H_2={P_02\over \rho^4}\ .
\ee

We now observe that both our metric \eq{hff} as well as its $\b \ra 0$
limit given in \eq{hff2} take the same form at spatial infinity reached
by taking the $\rho \ra \infty$ limit, and the resulting metric is

\be ds^2_6 = {4k^2\sqrt{{\widehat Q}_0}\over g_1^2P_02}\left(d\rho^2
+{g_1 P_0\over 8k}\,\rho^2\,\left(\s_1^2+\s_2^2 + {2g_1P_0\over
k}\,\s_3^2+ {2g_1\over k\,{\widehat Q}_0}\, dx^\mu dx_\mu\right)\right)\
,\la{cone}\ee

where ${\widehat Q}_0 \equiv \left({\widetilde Q}_0P_0-Q_0{\widetilde
P}_0\right)/P_0$. This metric indeed describes a cone over the product of
Minkowski$_2 \times$ the squashed 3-sphere \cite{Guven:2003uw}. In this
limit $F_7$ vanishes, $F_3$ and $H$ are finite but the dilaton diverges
as $e^\vf \ra {{\widehat Q}_0\over P_02}\,\rho^4$.

Our metric \eq{hff} has a horizon at $\rho=0$, just as its $\b \ra 0$
limit given in \eq{hff2} does. In this limit we obtain

\be ds^2_6=\sqrt{\b\over Q_0P_0}\,\left[\rho^2 dx^\mu dx_\mu
+{kQ_0P_0\over 2g_1\rho^2}\,\left(\s_1^2+\s_2^2+{2g_1P_0\over
k}\,e^{{-2\b P_02\over \rho^4}}\,\s_3^2\right) +{4k^2\b^2 P_0\over
g_1^2}\,e^{{-2\b P_0^2\over \rho^4}}\,{d\rho^2\over \rho^{12}} \right]\ .
\la{hor}\ee

Furthermore, while $H$ vanishes and $F_3,F_7$ become constants, the
dilaton diverges as $e^\vf \ra \b Q_0P_0/\rho^4$ in this limit.
Interestingly, taking the $v_0 \ra 0$ limit of this near horizon metric
\eq{hor} does not yield the same result as first taking such a limit in
the full metric \eq{hff} and then going to the horizon at $\rho=0$. In
the latter case, as shown in \cite{Guven:2003uw}, one obtains a direct
product of $AdS_3$ with squashed 3-sphere.


\section{Discussion}


In this paper we have given the precise form of the potential for the
scalars in the hypermatter multiplet of the only known anomaly free
gauged $(1,0)$ supergravity in $D=6$ in the absence of linear multiplets.
The model has the gauge group of $E_6\times E_7\times U(1)_R$. The
hyperscalars are charged with respect to $U(1)_R$ and transform in the
912 dimensional pseudo real representation of $E_7$. They are singlets of
$E_6$. We showed that the potential has a unique minimum at $\phi=0$.
Despite the fact that there is no obvious mass term in the D=6 action for
the scalars their Kaluza Klein tower, on a background of $R^4\times K_2$
will be all massive due to their $U(1)\times E_7$ charges. The
hypermatter fermions on the other hand will give rise to plenty of chiral
fermions in $D=4$. Thus opening the road for a detailed phenomenological
study of our model.

One interesting direction in further study of our model would be
cosmological investigation along the lines of \cite{Cline:2003ak,
Bringmann:2003pz,Bringmann:2003sz}. The quantum loop of the massive
scalars in this model are the natural candidates to stabilize the radius
of the compact space in a cosmological context at finite temperature. It
has been shown long time ago that the effect of such loops can generate a
constant radius for the internal space while the scale factor of our
3-dimensional universe expands according to the standard Friedman
Robertson Walker law \cite{Randjbar-Daemi:1983jz}. It is a very
interesting question to seek for an accelerating universe solution. Such
solution should exist according to the criteria given in
\cite{Townsend:2003qv}.

In this paper we also constructed a dyonic string solution in the same
model. Our solution leaves $1/4$ of the original supersymmetries, i.e.
one complex supersymmetry in $1+1$ dimensions, unbroken. Furthermore a
$U(1)$ component of the $E_7$ gauge field needs to be nonzero. In fact it
assumes a rather complicated form. The solution approaches a cone as
$r\rightarrow \infty$ over a squashed $S^3\times$ Minkowski$_2$, while at
$r=0$ it has a horizon.

Another important question is to complete the search, initiated in
\cite{Cvetic:2003xr},  for a higher dimensional origin of this gauged
supergravity model in $D=6$.

\bigskip


\section*{Acknowledgments}

\bigskip

We are very grateful to John Strathdee for his collaboration at the early
stages of this work. We thank Gary Gibbons, Rahmi G\"uven, Jim Liu, Hong
Lu, H. Nishino, Chris Pope and A. Sagnotti for useful discussions. E.S.
would like to thank the Abdus Salam International Center for Theoretical
Physics for hospitality. The work of E.S. is supported in part by NSF
grant PHY-0314712.

\newpage


\noindent{\Large\bf Appendix}\\

\noindent{\large\bf The Anomaly Polynomial}\\


There are few misprints in the anomaly polynomial formulae of
\cite{Randjbar-Daemi:wc} which we wish to correct here. We begin by
listing the individual contributions:

\bea P(\psi_\m) &=& \frac{5}{24} F_1^4 - \frac{19}{96} F_1^2 ~trR^2
+\frac{1}{5760}~\left[245~\tr~R^4
               -\frac{5\times 43}{4}~(\tr~R^2)^2 \right] \ ,
\la{p1}
\\&&\nn\\
-2P(\psi_R) &=& \frac1{24}~\Tr_{912}~F^4 +\frac1{96}~\Tr_{912}~F^2~\tr
R^2
    +\frac{912}{5760}~\left[\tr~R^4 +\frac54~(\tr R^2)^2\right]\ ,
\la{p2}
\\&&\nn\\
-P(\chi_R) &=& \frac{1}{24} F_1^4+ \frac{1}{96} F_1^2 ~trR^2 +
\frac{1}{5760}~\left[ \tr~R^4 +\frac{5}{4}~(\tr~R^2)^2\right] \ , \la{p3}
\\&&\nn\\
P(\l_L) &=& \frac{1}{24}(~\Tr_{78}~F^4+6~F_1^2~\Tr_{78}F^2+ 78F_1^4)
\nn\\&&\nn\\
&&
    +\frac{1}{24}(~\Tr_{133}~F^4 + 6~F_1^2~\Tr_{133}~F^2 +
    133F_1^4)+\frac{1}{24}~F_1^4
\nn\\&&\nn\\
&&
    +\frac1{96}~\left[\Tr_{78}~F^2 +\Tr_{133}~F^2
+(78+133+1)F_1^2\right]~\tr~R^2
\nn\\&&\nn\\
&&+\frac{(78+133+1)}{5760}~\left[\tr~R^4 +\frac54~(\tr R^2)^2\right]\ ,
\la{p4} \eea

where $F_1$ denotes the $U(1)_R$ field strength. Using the relations
between the traces in various representations involved above and the
fundamental representations, provided in \cite{Randjbar-Daemi:wc}, and
adding all the contributions, we find that
$P=P(\psi_\m)+P(\psi_R)+P_(\chi_R)+P(\l_R)$ is given by

\bea P &=& -\frac{1}{16}~(\tr R^2)^2 +\frac{1}{24}~\tr R^2~\Tr_{27}F^2
    -\frac{1}{8}~\tr R^2~\Tr_{56}F^2 +2~F_1^2~trR^2
\nn\\ && +\frac{1}{48}~\Tr_{27}~(F^2)^2+~F_1^2~Tr_{27}F^2
-\frac{3}{64}~(\Tr_{56}~F^2)^2 \nn\\ && +\frac{3}{4}~F_1^2~\Tr_{56}~F^2
+9F_1^4\ . \la{p} \eea

The signs of the $(\tr R^2)^2$ terms in \eq{p2}, \eq{p3}, \eq{p4} and a
factor of two in the coefficient of $\tr R^2~\Tr_{56}F^2 $ in \eq{p} have
been corrected relative to those in \cite{Randjbar-Daemi:wc}.

It is possible to factorize this expression and write it as

\bea P &=& -\frac{1}{16}\left(~tr R^2 +4 F_1 ^2 + \frac{1}{3}~Tr_{27}F^2
+ \frac{1}{2}~Tr_{56}F^2\right)\left(~tr R^2 -36 F_1 ^2 -~Tr_{27}F^2
+\frac{3}{2}~Tr_{56}F^2\right)\nn\\
&\equiv& -X_4 {\tilde X}_4\ ,\eea

where, upon writing the traces in the adjoint representations by means of
the formula given in \cite{Randjbar-Daemi:wc}, we have
\bea X_4&=& \frac14 \left(~v_L~tr R^2 + v_1~F_1^2 + ~v_6~Tr_{78} F^2
+~v_7~Tr_{133} F^2 \right )\nn\w2
 {\tilde X}_4&=& \frac14 \left(~{\tilde v}_L~tr R^2 + ~{\tilde v}_1~F_1^2
 +~{\tilde v}_6~Tr_{78} F^2
+~{\tilde v}_7~Tr_{133} F^2 \right )\ , \eea

where $(v_L,v_1,v_6,v_7)=(1,4,1/12,1/6)$ and $( {\tilde v}_L,{\tilde
v}_1,{\tilde v}_6,{\tilde v}_7)=(1,-36,-1/4,1/2)$.

\bigskip


\noindent{\bf The Einstein and Dilaton Field Equations }\\


Writing  the Einstein's equation for the model as $R_{rs}=S_{rs}$, where,
we recall that $r,s={\tilde 0}, {\tilde 1}, 1,2,3,4$ label the tangent
space frame defined in \eq{frame}, the non-vanishing components of
$R_{rs}$ evaluated for the ansatz \eq{an} are

\bea R_{\mu\nu} &=& -\left[ \fft{{c'}^2}{h^2\, c^2} + \fft{2 a'\, c'}{a\,
c\, h^2} + \fft{b'\, c'}{b\, c\, h^2}  + \fft1{c\, h}\,
\Big(\fft{c'}{h}\Big)' \right]\, \eta_{\mu\nu}\ ,\nn\w2
R_{11}&=& R_{22} = -\fft{2 a'\, c'}{a\, c\, h^2} - \fft{a'\, b'}{a\, b\,
h^2} - \fft{{a'}^2}{a^2\, h^2} - \fft1{a\, h}\,
\Big(\fft{a'}{h}\Big)'-\fft{b^2}{2a^4} + \fft1{a^2}\ ,\nn\w2
R_{33}&=& -\fft{2 b'\, c'}{a\, c\, h^2} - \fft{2 a'\, b'}{a\, b\, h^2}
-\fft1{b\, h}\, \Big( \fft{b'}{h}\Big)'
  +\fft{b^2}{2a^4}\ ,\nn\w2
R_{44} &=& -\fft{2}{a\, h}\, \Big(\fft{a'}{h}\Big)' - \fft{1}{b\, h}\,
\Big(\fft{b'}{h}\Big)' - \fft{2}{c\, h}\, \Big(\fft{c'}{h}\Big)'\ , \eea

while the non-vanishing components of $S_{rs}$ take the form

\bea
S_{\mu\nu}&=& -\left[ \fft{k^2}{8 a^4}\, e^{\ft12\vf}\,
+\ft18\Big({v^2\over a^4}+{v'^2\over h^2b^2}\Big)e^{\ft12\vf}\,+ \ft14 \,
\Big( \fft{u^2}{h^2\, c^4} + \fft{P^2}{a^4\, b^2}\Big)\,e^\vf - \ft12
g_1^2\, e^{-\ft12\vf}\right]\,\eta_{\mu\nu}\ ,\nn\w2
S_{11}&=&S_{22}= \fft{3k^2}{8 a^4}\, e^{\ft12\vf}\, + \Big({3v^2\over
8a^4}-{v'^2\over 8h^2b^2}\Big)e^{\ft12\vf}\,+ \ft14\, \Big(
\fft{u^2}{h^2\, c^4} + \fft{P^2}{a^4\, b^2}\Big)\, e^\vf\, + \ft12
g_1^2\, e^{-\ft12\vf}\ ,\nn\w2
S_{33}&=& -\fft{k^2}{8 a^4}\, e^{\ft12\vf}\, - \Big({v^2\over
8a^4}-{3v'^2\over 8h^2b^2}\Big)e^{\ft12\vf}\,+ \ft14 \, \Big(
\fft{u^2}{h^2\, c^4} + \fft{P^2}{a^4\, b^2}\Big)\,e^\vf\, +\ft12 g_1^2\,
e^{-\ft12\vf}\ ,\nn\w2
S_{44}&=& -\fft{k^2}{8 a^4}\, e^{\ft12\vf}\, - \Big({v^2\over
8a^4}-{3v'^2\over 8h^2b^2}\Big)e^{\ft12\vf}\, - \ft14\, \Big(
\fft{u^2}{h^2\, c^4} + \fft{P^2}{a^4\, b^2}\Big)\, e^\vf\, +\ft12 g_1^2\,
e^{-\ft12\vf}\ .\la{t} \eea

Finally, writing the dilaton field equation as ${\square\, \vf} = J$, the
evaluation of $J$ for our the ansatz \eq{an} yields

\be J=\fft{k^2}{2 a^4}\, e^{\ft12\vf}\, +\ft12\Big({v^2\over
a^4}+{v'^2\over h^2b^2}\Big)e^{\ft12\vf}\,+ \Big( \fft{P^2}{a^4\,
b^2}-\fft{u^2}{h^2\, c^4} \Big)\,e^\vf - 2 g_1^2\, e^{-\ft12\vf}\ .\ee

\newpage


\ed
\begin{thebibliography}{99}



\bibitem{Salam:1984cj}
A.~Salam and E.~Sezgin,
   Phys.\ Lett.\ B {\bf 147}, 47 (1984).


\bibitem{Gibbons:2003di}
G.~W.~Gibbons, R.~G\"uven and C.~N.~Pope,
arXiv:hep-th/0307238.


\bibitem{Randjbar-Daemi:wc}
S.~Randjbar-Daemi, A.~Salam, E.~Sezgin and J.~Strathdee,
Phys.\ Lett.\ B {\bf 151}, 351 (1985).


\bibitem{Nishino:dc}
H.~Nishino and E.~Sezgin,
Nucl.\ Phys.\ B {\bf 278}, 353 (1986).


\bibitem{Nishino:1997ff}
H.~Nishino and E.~Sezgin,
Nucl.\ Phys.\ B {\bf 505}, 497 (1997), arXiv:hep-th/9703075.


\bibitem{Berkooz:1996iz}
M.~Berkooz, R.~G.~Leigh, J.~Polchinski, J.~H.~Schwarz, N.~Seiberg
and E.~Witten,
Nucl.\ Phys.\ B {\bf 475}, 115 (1996), arXiv:hep-th/9605184.


\bibitem{Cvetic:2003xr}
M.~Cvetic, G.~W.~Gibbons and C.~N.~Pope,
arXiv:hep-th/0308026.


\bibitem{Gherghetta:2002nq}
T.~Gherghetta and A.~Kehagias,
Phys.\ Rev.\ D {\bf 68}, 065019 (2003), arXiv:hep-th/0212060.


\bibitem{Aghababaie:2003wz}
Y.~Aghababaie, C.~P.~Burgess, S.~L.~Parameswaran and F.~Quevedo,
arXiv:hep-th/0304256.

\bibitem{Randjbar-Daemi:1982hi}
S.~Randjbar-Daemi, A.~Salam and J.~Strathdee,
Nucl.\ Phys.\ B {\bf 214}, 491 (1983).


\bibitem{Carroll:2003db}
S.~M.~Carroll and M.~M.~Guica,
arXiv:hep-th/0302067.


\bibitem{Navarro:2003vw}
I.~Navarro,
JCAP {\bf 0309}, 004 (2003), arXiv:hep-th/0302129.
Class.\ Quant.\ Grav.\  {\bf 20}, 3603 (2003), arXiv:hep-th/0305014.


\bibitem{Dvali:2002pe}
G.~Dvali, G.~Gabadadze and M.~Shifman,
Phys.\ Rev.\ D {\bf 67}, 044020 (2003), arXiv:hep-th/0202174.

\bibitem{Guven:2003uw}
R.~G\"uven, J.~T.~Liu, C.~N.~Pope and E.~Sezgin,
arXiv:hep-th/0306201.


\bibitem{Duff:1995yh}
M.~J.~Duff, S.~Ferrara, R.~R.~Khuri and J.~Rahmfeld,
Phys.\ Lett.\ B {\bf 356}, 479 (1995) [arXiv:hep-th/9506057].


\bibitem{Duff:1996cf}
M.~J.~Duff, H.~Lu and C.~N.~Pope,
Phys.\ Lett.\ B {\bf 378}, 101 (1996) [arXiv:hep-th/9603037].


\bibitem{Percacci:1998ag}
R.~Percacci and E.~Sezgin,
arXiv:hep-th/9810183.

\bibitem{Randjbar-Daemi:1983bw}
S.~Randjbar-Daemi, A.~Salam and J.~Strathdee,
Phys.\ Lett.\ B {\bf 124}, 345 (1983) [Erratum-ibid.\ B {\bf 144}, 455
(1984)].

\bibitem{Dvali:2001qr}
G.~R.~Dvali, S.~Randjbar-Daemi and R.~Tabbash,
Phys.\ Rev.\ D {\bf 65} (2002) 064021, arXiv:hep-ph/0102307.


\bibitem{Cline:2003ak}
J.~M.~Cline, J.~Descheneau, M.~Giovannini and J.~Vinet,
JHEP {\bf 0306}, 048 (2003), arXiv:hep-th/0304147.

\bibitem{Bringmann:2003pz}
T.~Bringmann and M.~Eriksson,
JCAP {\bf 0310}, 006 (2003), arXiv:astro-ph/0308498.


\bibitem{Bringmann:2003sz}
T.~Bringmann, M.~Eriksson and M.~Gustafsson,
Phys.\ Rev.\ D {\bf 68}, 063516 (2003), arXiv:astro-ph/0303497.

\bibitem{Randjbar-Daemi:1983jz}
S.~Randjbar-Daemi, A.~Salam and J.~Strathdee,
Phys.\ Lett.\ B {\bf 135}, 388 (1984).

\bibitem{Townsend:2003qv}
P.~K.~Townsend,
arXiv:hep-th/0308149.



\end{thebibliography}
